\documentclass{article}[12pt]
\usepackage{amsmath}
\usepackage{amssymb}
\usepackage{amsthm}
\usepackage{amsfonts}
\usepackage[dvips]{graphicx}
\usepackage{color}

\def\abstract#1{\vskip 7mm
        \begin{center}{\large Abstract}\par \smallskip
                \begin{minipage}[c]{12cm}
                        \small #1
                \end{minipage}
        \end{center}
}

\def\title#1{\begin{center}{\Large\bf #1}\end{center}}
\def\author#1{\vskip 5mm \begin{center}{#1}\end{center}}
\def\address#1{\begin{center}{\it #1}\end{center}}
\newcommand{\ssmatrix}[4]%
{\begin{pmatrix} #1 & #2 \\ #3 & #4 \end{pmatrix}}
\makeatletter
\@ifundefined{lesssim}{}{}
\@ifundefined{gtrsim}{}{}
\def\vereq#1#2{\lower3pt\vbox{\baselineskip1.5pt \lineskip1.5pt
\ialign{$\m@th#1\hfill##\hfil$\crcr#2\crcr\sim\crcr}}}
\makeatother

    {\mbox{}\hfill\mbox{\rm\normalsize$\Box$}\\%
    }
\usepackage{color}
\input{colordvi.tex}
\newcommand*{\D}{{\rm d}}
\begin{document}

\title{Teichm\"{u}ller parameters for multiple BTZ black hole spacetime}

\author{Yasunari Kurita}

\address{Kanagawa Institute of Technology, Atsugi, Kanagawa 243-0292, Japan}

\author{Masaru Siino}

\address{Department of Physics, Tokyo Institute of Technology, Tokyo 152-8551, Japan}

\abstract{
We investigate the Teichm\"{u}ller parameters for a 
Euclidean multiple BTZ black hole spacetime.
To induce a complex structure in the asymptotic boundary of such a spacetime, we consider the limit in which two black holes are at a large distance from each other.
In this limit, we can approximately determine the period matrix $\Omega_{ij}$ (i.e., the Teichm\"{u}ller parameters) for the spacetime boundary 
by using a pinching parameter.
The Teichm\"{u}ller parameters are essential 
in describing the partition function for the boundary conformal field theory (CFT).
We provide an interpretation of the partition function for the genus two extremal 
boundary CFT proposed by Gaiotto and 
Yin that it is relevant to double BTZ black hole spacetime.
}
%\pacs{}
%\preprint{}
%\maketitle

\section{Introduction}

Black holes are fascinating objects for investigating quantum aspects of gravity, and
their thermodynamic properties provide a key to understanding quantum gravity.
For example, in the 1970s, the Euclidean path integral approach to quantum gravity revealed that black hole thermodynamics can be formulated using a Euclidean partition function, similar to the case for normal statistical mechanics~\cite{Gibbons:1976ue,Hawking:1979ig}.
By the 1990s, black hole thermodynamics had become a major approach to understanding gravity in the context of string theory. 
One major breakthrough in string theory was the derivation of 
the Bekenstein-Hawking entropy by counting black hole microstates, which gives a precise statistical mechanical interpretation of black hole entropy~
\cite{Strominger:1996sh}.

Maldacena discovered a relation between a $d+1$ dimensional gravitational system described in terms of the weak limit of string theory and a conformal field theory on the $d$ dimensional boundary
\cite{Maldacena:1997re,Gubser:1998bc,Witten:1998qj,Aharony:1999ti}.
This relation is referred to as the anti-de Sitter/conformal field theory (AdS/CFT) correspondence, and it has been applied in a very wide range of research fields, even in theoretical condensed matter physics.
After the discovery of AdS/CFT, AdS black holes have been growing in importance.

The simplest type of asymptotically AdS black hole is a BTZ black hole~\cite{Banados:1992wn,Banados:1992gq}.
The existence of a black hole even in three dimensions makes three-dimensional (3D) gravity much more interesting. 
A BTZ black hole has a horizon with a positive length and a corresponding Bekenstein-Hawking entropy.
This might be a good chance for constructing a solvable model with quantum black holes~\cite{Witten:2007kt}.
A generalization of a BTZ black hole spacetime was presented by Brill~\cite{Brill:1995jv}, who considered a spacetime including multiple BTZ black holes.
The multiple BTZ black hole geometries are solutions to 3D AdS gravity, and should therefore be relevant to the AdS/CFT correspondence. 
However, to our knowledge, no description of multiple BTZ black holes in terms of dual CFT has yet been presented.
Furthermore, there have been no studies on black hole thermodynamics for a multiple BTZ black hole 
spacetime.

The dual CFT lives on the boundary of an asymptotically AdS spacetime,
and it is important to understand the complex structure (conformal structure) of the 
boundary at infinity.
In fact, the moduli parameters for the boundary parameterize 
the partition functions of the boundary CFT and also include information 
on the global AdS geometry, and thus play an essential role in AdS/CFT.
In this sense, it is meaningful to investigate the Teichm\"{u}ller (moduli) parameters 
for the boundary of a multiple black hole geometry.

In this article, we express the Teichm\"{u}ller parameters for the boundary
of a Euclidean multiple BTZ black hole geometry using a period matrix for the case of a double BTZ black hole spacetime.
In the next section, as an exercise, we first consider a single BTZ black hole 
and calculate the Teichm\"{u}ller parameter for the boundary torus.
We provide a short review showing that a Euclidean BTZ black hole can be obtained as a quotient space of a 3D hyperbolic space $\bf{H_3}$ by a discrete subgroup $\Gamma$ of its isometry group SO(3,1).
In section \ref{sec:multiBTZ}, 
we construct a Euclidean multiple BTZ black hole geometry that is also a quotient space of $\bf{H_3}$, in a manner essentially equivalent to that used by Brill.
This is achieved by considering other generators of $\Gamma$ for the Euclidean single BTZ black hole spacetime.
We demonstrate that the boundary of the Euclidean multiple BTZ black hole spacetime is a Riemann surface with a genus $g\geq 2$, which allows a conformal structure.
In section \ref{sec:teich-multiBTZ}, we derive the period matrix (Teichm\"{u}ller parameters) 
as a power series in the pinching parameter for the boundary of a Euclidean double BTZ black hole spacetime in which the two black holes are assumed to be separated by a large distance. 
We also provide a physical interpretation of the pinching parameter 
in terms of the relative positions and orientations of the two BTZ black holes. 

As an application of the Teichm\"{u}ller parameters, 
we consider the corresponding CFT for the double BTZ black hole spacetime in section \ref{sec:partition-function}.
The CFT should live on a $g= 2$ Riemann surface at the boundary. Witten~\cite{Witten:2007kt} argued that the dual CFT corresponding to pure 3D AdS gravity is an extremal CFT (ECFT).
The genus two modular invariant partition function for ECFTs was calculated by Gaiotto and Yin
for several values of the central charge~\cite{GaiottoYin},
and we suggest that the modular invariant genus two partition function for an ECFT includes the contribution of the double BTZ black hole geometry.
Finally, we discuss some physical properties of double BTZ black holes based on the 
genus two partition function in section \ref{sec:discussion}.

\section{Complex structure of boundary of single BTZ black hole}
\label{sec:singleBTZ}

As is well known, the BTZ black hole spacetime is identified to the 
quotient spacetime of the 3D anti de Sitter spacetime by a discrete subgroup 
of its isometry group SO(2,2), which is generated by a Lorentz boost 
and a rotational transformation~\cite{Banados:1992gq}.
First of all, we prepare a single BTZ black hole with a Euclidean signature 
and induce complex structure at its boundary.

The metric function for the Euclidean BTZ black hole is given by Wick 
rotation of a Lorentzian BTZ black hole, 
\begin{align}
\D s^2= \frac{(r^2-r_+^2)(r^2-r_-^2)}{l^2r^2}\D t^2+\frac{l^2r^2\D 
 r^2}{(r^2-r_+^2)(r^2-r_-^2)}+r^2\left(\D 
 \phi - i\frac{r_+r_-}{lr^2}\D t \right)^2,
\label{eqn:BTZ}
\end{align}
where $l$ is the radius of curvature, and $r_+$ and $r_-$ are the original 
horizon radii in the Lorentzian metric, and are related to the black hole mass $M$ and 
the angular momentum $J$ as 
\begin{align}
r_+^2+r_-^2=Ml^2,
\label{eq:mass-horizon-radius}
\end{align}
and
\begin{align}
\frac{ir_+r_-}{l}=  \frac{J_E}{2}= -i \frac{J}{2}. 
\label{eq:angular-momentum-horizon-radius}
\end{align}
The angular momentum $J$ is analytically continued to the Euclidean $J_E=-iJ$ since the spacetime changes its signature.
In the Wick rotation, Lorentzian time is replaced by $-it$ (in 
this paper, $t$ represents Euclidean time) and $r_-$ becomes a pure imaginary number.

Since a Euclidean 3D vacuum spacetime with a negative cosmological constant is locally equivalent to a 3D hyperbolic space, which is the Euclidean version of anti de Sitter spacetime, these coordinates are related to a pseudosphere in a flat $E_{3,1}$ space as follows.
The pseudosphere is given by
\begin{align}
-x_0^2+x_1^2+x_2^2+x_3^2=-l^2=B-A,
\end{align}
where
\begin{align}
A=\frac{r^2-r_-^2}{r_+^2-r_-^2}l^2, \ \ B=\frac{r^2-r_+^2}{r_+^2-r_-^2}l^2.
\label{eq:ads}
\end{align}

The BTZ black hole is embedded in the flat $E_{3,1}$ space by 
the following coordinate parameterization:
\begin{align}
x_0&=\sqrt{A}\cosh \tilde{\phi},
\label{coord:pseudosphere-BTZ1} \\
x_1&=\sqrt{B}\cos \tilde{t},
\label{coord:pseudosphere-BTZ2} \\
x_2&=\sqrt{B}\sin \tilde{t},
\label{coord:pseudosphere-BTZ3} \\
x_3&=-\sqrt{A}\sinh \tilde{\phi},
\label{coord:pseudosphere-BTZ4}
\end{align}
where the BTZ coordinates $(t,\phi)$ are related to $(\tilde{t},\tilde{\phi})$ by
\[
\tilde{\phi}=\frac1l \left( -\frac{ir_- t}l  +r_+ \phi\right),\ \ \ 
\tilde{t}=\frac1l \left(\frac{r_+ t}l + i r_- \phi\right).
\]
In this parameterization, the pseudosphere is fully covered in the range $r>r_+, -\infty<\tilde{\phi}<\infty, -\pi<\tilde{t}<\pi$. 
Here, $\tilde{t}$ possesses Euclidean time periodicity as usual. 
We obtain a Euclidean BTZ metric (\ref{eqn:BTZ}) as an induced metric on 
the pseudosphere in $E_{3,1}$ with $\D s^2=-\D x_0^2+\D x_1^2+\D x_2^2+\D x_3^2$.
It should be noted that the range of $\phi$ is
$-\infty<\phi<\infty$. 
If the metric (\ref{eqn:BTZ}) is regarded as that of the black hole spacetime, 
the angular coordinate should have a period $\phi \sim \phi +2\pi$.
Therefore, the spacetime has to be identified along the Killing vector
\begin{align}
\xi_{\phi}:=\frac{\partial}{\partial \phi}=
-\frac{r_+}{l} J_{03}-\frac{r_-}{ l} J_{12},
\label{eq:ki1}
\end{align}
where  $J_{03}:=x_3\partial_0+x_0\partial_3$ and 
$J_{12}:=ix_2\partial_1-ix_1\partial_2$, i.e., $J_{ij}$ is an infinitesimal generator of SO(3,1).
The identification is given by an exponential map along $\xi_{\phi}$ 
\begin{align}
\gamma=\exp\left(2\pi \xi_{\phi} \right)
=\exp\left[ - \frac{2\pi r_+}l J_{03}
        - \frac{2\pi r_-}lJ_{12}\right],
\label{eqn:exp}
\end{align}
which generates a discrete subgroup $\Gamma=\{ \gamma^i | i\in \mathbb{Z} \}\sim {\bf Z}$ of isometry SO(3,1).
Since $J_{03}$ and $J_{12}$ commute, this identification involves a Lorentz boost in the $x_3$ direction (the direction of $\tilde{\phi}$) 
with a boost angle of $2\pi r_+/l$, and a rotation in the ($x_1,x_2$)-plane
(in the direction of Euclidean time $\tilde{t}$) 
of $2\pi \omega/l$, 
where we have defined $\omega=ir_-=\frac{J_El}{2r_+}$. 

In order to realize complex structure in the hyperbolic space, 
it is convenient to introduce the Poincar\'{e} coordinates ($X,Y,Z$). 
This coordinate system is related to the pseudosphere in $E_{3,1}$ by the parameterization
\begin{align}
X:=\frac{x_1}{x_0+x_3}, \quad
Y:=\frac{x_2}{x_0+x_3}, \quad 
U:=\frac{l}{x_0+x_3}.
\end{align}
Therefore, its metric function is given by
\begin{align}
\tilde{\D s^2}=l^2\frac{\D X^2+\D Y^2+\D U^2}{U^2},
\end{align}
which may be recognized as the standard metric for the upper half-space 
model of 3D hyperbolic space $\bf H_3$. 
In this coordinate system, the hyperbolic space is mapped onto the upper 
half ($U>0$) of $\bf R^3$ and its boundary at infinity 
($x_1^2+x_2^2+x_3^2\sim \infty$) corresponds to the plane $U=0$
\cite{Aharony:1999ti,Bayona:2005nq}.

The boost generator in (\ref{eqn:exp}) can then be expressed in Poincar\'{e} coordinates as
\begin{align}
-\frac{\pi r_+}lJ_{03}=\frac{\pi r_+}l(U\partial_U+X\partial_X+Y\partial_Y),
\label{eq:dilatation-generator}
\end{align}
and an exponential map of (\ref{eq:dilatation-generator}) gives a homothetic expansion in $(X,Y,U)$ coordinates by $\exp(2\pi r_+/l)$.
The rotation in the $(x_1,x_2)$ plane in (\ref{eqn:exp}) is identified 
to a rotation in the $(X,Y)$ plane with an angle $2\pi \omega/l$.

\begin{figure}[tp]
\centering
\includegraphics[width=9cm,clip]{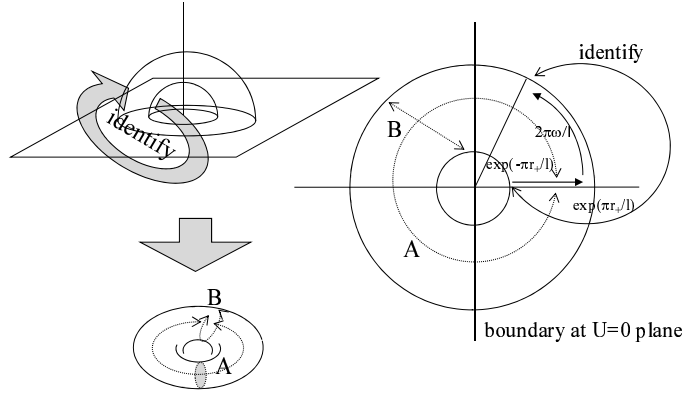}
\caption{Identification that produces a solid torus for the boundary defined by the $U=0$
plane (right) and the upper half-space (left).
In the left figure, the Euclidean black hole is obtained by identifying the inner and outer 
 hemisphere with a twist around the $U$ axis.
In the right figure, $A$ and $B$ represent the canonical homology basis on the boundary.}
\label{fig:torus}
\end{figure}

Equivalently, this identification can be easily seen by
relating the Poincar\'{e} coordinates to the Euclidean BTZ coordinates~\cite{Carlip:1994gc},
\begin{eqnarray}
X &=& \left(\frac{r^2-r_+^2}{r^2-r_-^2}\right)^{1/2}\cos
\left(
\frac{r_+}{l^2}t+\frac{\omega}{l}\phi
\right) 
\exp\left[\frac{r_+}{l}\phi-\frac{\omega}{l^2}t \right], \\
Y &=&  \left(\frac{r^2-r_+^2}{r^2-r_-^2}\right)^{1/2}\sin
\left(
\frac{r_+}{l^2}t+\frac{\omega}{l}\phi
\right) 
\exp\left[\frac{r_+}{l}\phi-\frac{\omega}{l^2}t \right],   \\
U &=& \left(\frac{r_+^2-r_-^2}{r^2-r_-^2}  \right)^{1/2}
\exp\left[ \frac{r_+}{l}\phi-\frac{\omega}{l^2}t  \right].
\end{eqnarray}
We now introduce the ``spherical'' coordinates
\begin{eqnarray}
X &=& R\cos\tilde{t} \cos\chi, \\
Y &=& R\sin\tilde{t} \cos\chi, \\
U &=& R\sin \chi, \qquad (\chi>0).
\end{eqnarray}
The identification, $\phi \sim \phi+2\pi$, then becomes
\begin{eqnarray}
(R,\ \tilde{t},\ \chi)\sim (Re^{\frac{2\pi r_+}{l}},\ 
 \tilde{t}+\frac{2\pi\omega}{l},\ \chi),
\end{eqnarray}
which means that the Euclidean BTZ black hole is obtained by identifying two 
hemispheres with a twist around the $U$ axis, as shown in Figure~\ref{fig:torus}. 
The ratio of the radius of these two hemispheres is 
$e^{\frac{2\pi r_+}{l}}$ and the twist angle is $2\pi \omega/l$.
Topologically, the resulting manifold is a solid torus and its boundary 
is a torus, as shown in Figure~\ref{fig:torus}.

Complex structure is then naturally introduced into the hyperbolic space by the quaternion 
$Q=Z+jU$ with $Z=X+iY$, 
indicating the complex structure on the boundary at infinity. 
It is consistent that the isometry group SO(3,1) in Euclidean $\rm{AdS}_3$ 
(identical to ${\bf H_3}$) is locally isomorphic to SL(2,${\mathbb C}$), 
which is an automorphism of the upper half space $(X,Y,U)$ of ${\bf H_3}$ 
including the rotation and expansion.
Using the quaternion, the identification on the boundary to give a 
Riemann surface can be expressed as $Q'=e^{i2\pi \omega/l}e^{2\pi r_+/l}Q$,
which is an element of SL(2,$\mathbb{C}$).

For the Riemann surface on the boundary $U=0$ plane, a holomorphic 1-form is given by $\nu=\D Z/2\pi i Z$. 
The Teichm\"{u}ller parameter is then determined as follows:
\begin{align}
\int_A \nu&=1, \\
\int_B \nu&= \frac{\omega}{l}+i\frac{r_+}{l}= \tau_{\rm BTZ},
\label{eqn:teich}
\end{align}
where $A$ and $B$ represent the canonical homology basis shown in Figure~\ref{fig:torus}.
The parameter $\tau_{\rm BTZ}$ is a moduli parameter for the 
boundary torus of the Euclidean BTZ black hole.
However, since it is usual to use $\tau=-1/\tau_{\rm BTZ}$ 
instead of $\tau_{\rm BTZ}$, especially when describing the partition 
function for the boundary CFT~\cite{Maloney:2007ud}, 
we introduce 
\begin{align}
\tau = - \frac{1}{\tau_{\rm BTZ}} 
     = \frac{\omega l}{r_+^2-r_-^2}+i\frac{r_+ l}{r_+^2-r_-^2},
\label{eq:modular-transformation-AdS-BTZ}
\end{align}
which is a moduli parameter for the boundary torus for thermal AdS$_3$~\cite{Maldacena:1998bw,Brotz:1999xx}.
It should be noted that the boundary torus for the Euclidean BTZ spacetime and 
that for thermal AdS$_3$ are related by the modular transformation $\tau \to -1/\tau$ 
~\cite{Maldacena:1998bw,Kurita:2004yn}.

The moduli parameter $\tau$ can be expressed in terms of the period of the 
Euclidean time $\beta$ (or the inverse of the Hawking temperature $T_H$) 
and the Euclidean angular velocity $\Omega_E$ of the black hole 
\begin{eqnarray}
\beta=T_H^{-1}= \frac{2\pi r_+ l^2}{r_+^2-r_-^2},
\quad 
\Omega_E = \frac{\omega}{r_+ l},
\end{eqnarray}
as
\begin{eqnarray}
\tau = \frac{\beta}{2\pi}\left(\Omega_E+i \frac{1}{l}\right).
\label{eq:tau-beta-Omega}
\end{eqnarray}
Equation (\ref{eq:tau-beta-Omega}) is a well-known expression for the moduli 
parameter for the boundary torus~\cite{Brotz:1999xx,Banados:1998ta}.

\section{Euclidean multiple BTZ black hole spacetime and its boundary}
\label{sec:multiBTZ}

A method for constructing a multiple BTZ black hole geometry 
was described by Brill~\cite{Brill:1995jv}.
Since the BTZ black hole is simply a quotient spacetime of $\rm{AdS}_3$, it 
is possible to add BTZ black holes into the BTZ black hole geometry 
by considering further identifications by another discrete subgroup $\Gamma'$ of 
the isometry.
In this section, we construct Euclidean multiple BTZ black hole geometries 
using $\Gamma'$, which is essentially equivalent to the method described by Brill.\footnote{
To visualize the construction, Brill considered a double covered space.
The resultant manifold then possesses an extra $\bf{Z_2}$-symmetry.}

To construct the Euclidean multiple BTZ black hole geometry, 
the Poincar\'{e} disk model for 3D hyperbolic space ${\bf H}_3$ is convenient.
This corresponds to a 3D disk with radius of unity, accompanied by the metric
\begin{align}
\D s^2
= \frac{4l^2(\D x_p^2+\D y_p^2+\D z_p^2)}{(1-x_p^2-y_p^2-z_p^2)^2}.
\end{align}
The coordinate system $(x_p,y_p,z_p)$ is 
related to that for the pseudosphere in $E_{3,1}$ by
\begin{align}
x_p&=\frac{x_1}{l+x_0} ,
\label{coord:pseudosphere-disk1}\\
y_p&=\frac{x_2}{l+x_0} ,\label{coord:pseudosphere-disk2}\\
z_p&=\frac{x_3}{l+x_0} .\label{coord:pseudosphere-disk3}
\end{align}
The coordinate range is $x_p^2+y_p^2+z_p^2=r_p^2<1$ 
and the sphere $r_p=1$ is the boundary of the manifold at infinity.

\begin{figure}[tp]
\centering
\includegraphics[width=4cm,clip]{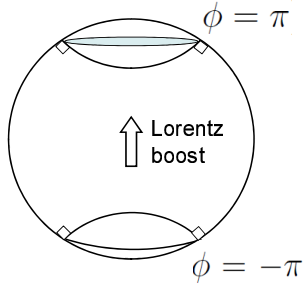}
\caption{In the disk model, a totally geodesic surface ($\phi=-\pi$) is 
mapped onto another totally
 geodesic surface ($\phi=\pi$) by an isometry corresponding to the Lorentz 
 boost generated by $J_{03}$.}
\label{fig:Lorentz-z}
\end{figure}

As discussed in section 2, the 
Euclidean BTZ black hole spacetime is a quotient spacetime of {\bf H}$_3$
produced using an 
identification by a discrete subgroup $\Gamma$ of the isometry of {\bf H}$_3$. 
More concretely, it is constructed by an identification between 
two parallel totally geodesic surfaces (corresponding to $\phi$ constant surfaces 
in the BTZ metric) which are mapped onto each other by a Lorentz boost along an axis and 
a twist about the axis.
In other words, the periodicity in the angular coordinate $\phi$ 
produces an identification 
between the two totally geodesic surfaces.
In the disk model, the totally geodesic surfaces are represented by spheres 
that meet the boundary sphere $r_p=1$ orthogonally.

\begin{figure}[tp]
\centering
\includegraphics[width=10cm,clip]{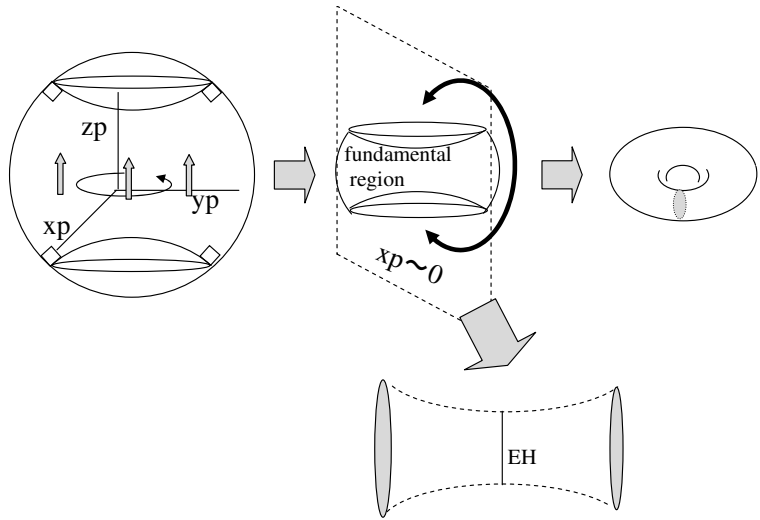}
\caption{The left figure shows a 3D disk with two totally 
 geodesic surfaces. 
The surface $\phi=-\pi$ (the lower surface crossing the boundary sphere orthogonally) 
is mapped onto the surface $\phi=\pi$ (the upper surface) 
by transformation along the $z_p$ axis 
and a rotation in the $(x_p,y_p)$ plane.
The identification between these two surfaces gives a Euclidean BTZ black 
 hole spacetime.
After the identification, the manifold becomes a solid torus with a 
 boundary at infinity, as shown in the right figure.
Taking a spatial slice, for example at $x_p \sim 0$, 
yields two boundaries at infinity connected by a single throat.}
\label{fig:E-BTZ}
\end{figure}

We now show how the isometry given by (\ref{eq:ki1}) appears in the disk model.
It should be noted that 
rotations around the coordinate axes $x_1, x_2$ and $x_3$ are 
simply rotations around the axes $x_p, y_p$ and $z_p$, respectively, 
and Lorentz boosts along the coordinate axes $x_1, x_2$ and $x_3$
correspond to isometries in the $x_p, y_p$ and $z_p$ directions, respectively. 
For example, a Lorentz boost along the $x_3$ axis generated 
by $J_{03}$ moves a totally geodesic surface 
to another totally geodesic surface in the $z_p$ direction, 
as shown in Figure~\ref{fig:Lorentz-z}.
We refer to the isometries in the $x_p, y_p$ and $z_p$ directions as 
$x_p, y_p$ and $z_p$ Lorentz boosts, respectively. 
Therefore, the isometry (\ref{eq:ki1}) involves a 
$z_p$ Lorentz boost with a boost angle $2\pi r_+/l$ 
and a rotation of $2\pi \omega/l$ around the $z_p$ axis 
in the direction of $\tilde{t}$.
We choose the totally geodesic surfaces $\phi=\pm\pi$ as 
the surfaces identified to construct the Euclidean BTZ black hole.
The mass $M$ and angular momentum $J$ of the black hole are
determined by the boost and twist angle in the identification, through
the relations (\ref{eq:mass-horizon-radius}) and (\ref{eq:angular-momentum-horizon-radius}). 
The Euclidean BTZ black hole spacetime 
is then composed of a part of the Poincar\'{e} disk bounded by 
two totally geodesic surfaces, as shown in Figure~\ref{fig:E-BTZ}. 
For a given spatial slice, for example at $x_p\sim 0$, 
two boundaries at infinity are connected by a single throat after the identification, 
and the resultant 3D manifold is a solid torus with a boundary at infinity.

\begin{figure}[tp]
\centering
\includegraphics[width=10cm,clip]{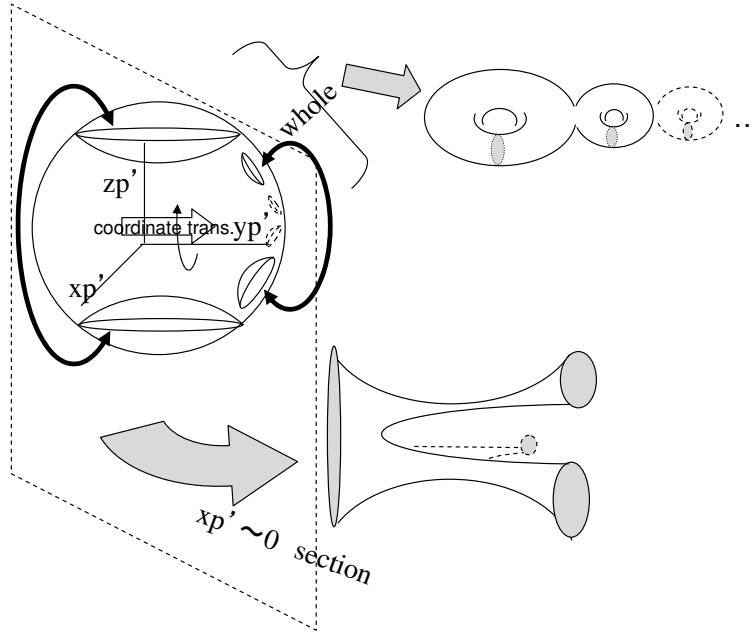}
\caption{Carrying out two identifications in {\bf H}$_3$ gives 
a Euclidean double BTZ black hole spacetime.
The manifold has the topology of a genus two solid torus.
A given spatial slice yields three boundaries at infinity connected by two throats.
The dashed lines indicate the case of adding a further BTZ black hole.}
\label{fig:E2-BTZ}
\end{figure}

We now construct a Euclidean spacetime including several BTZ black holes by making
further identifications.
We consider a coordinate transformation generated by 
a Lorentz boost and rotation. 
One may consider a $y_p$ Lorentz boost and a rotation around the $y_p$ axis 
without loss of generality.
We use boost and rotation angles of $2\alpha$ and $2\theta$, respectively, 
for later convenience. 
The parameter $2\theta$ allows a difference in orientation to exist between the two
black holes. 
These parameters will be used in section \ref{sec:epsilon}.
By this coordinate transformation, the totally geodesic surfaces which 
have been identified to construct the first BTZ black hole
move so as not to cross the $\phi=\pm\pi$ surfaces 
in the new coordinate system.\footnote{In order to obtain a smooth manifold, the 
boost angle should be larger than a certain value. Since we consider a 
sufficiently large boost angle in the present study, however, we do not discuss its minimum value.}
After the coordinate transformation, 
a new identification between the $\phi=\pm\pi$ surfaces can be considered, 
as illustrated in the left side of figure~\ref{fig:E2-BTZ}.
Different boost and rotation angles can be chosen to 
those used in the first identification.
To add a BTZ black hole with a mass $M'$ and an angular momentum $J'$ to 
the manifold,
the boost and rotation angles should be $2\pi r_+(M',J')/l$ and $2\pi\omega(M',J')$,
respectively, where $r_+(M',J')$ and $\omega(M', J')$ 
are determined by relations similar to 
(\ref{eq:mass-horizon-radius}) 
and (\ref{eq:angular-momentum-horizon-radius}).\footnote{
In Brill's construction of a double BTZ black hole geometry, 
the case of $J=0$ BTZ was suggested because of the $\bf{Z_2}$-symmetry appearing as a 
byproduct of the double cover.
On the other hand, we consider a more general situation without $\bf{Z_2}$-symmetry
and the case of $J\ne0$ double BTZ black hole geometry can be constructed.}
By this second identification, an additional BTZ black hole 
is brought into the manifold, and we refer to this geometry as E2-BTZ. The identifications generate a discrete subgroup $\Gamma'=\langle \gamma_1,\gamma_2\rangle$ of the isometry SO(3,1).
For a given spatial slice of the resultant manifold, 
for example at $x_p\sim0$, 
three boundaries at infinity are connected by two throats. 
The overall Euclidean spacetime is a solid double torus with a boundary at 
infinity, as illustrated in Figure~\ref{fig:E2-BTZ}.
Therefore, the topology of the boundary is a double torus.

By making further similar identifications, 
a Euclidean multiple BTZ black hole spacetime can be constructed, 
as discussed by Brill~\cite{Brill:1995jv}.
If the spacetime includes $g$ Euclidean BTZ black holes,
it is topologically a genus $g$ solid torus 
and its boundary is a genus $g$ torus.
\footnote{The number of BTZ black holes, $g$, refers to the number of throat structures 
(identical to the number of identifications) 
while, in \cite{Brill:1995jv}, the number of black holes indicates the number of disconnected components of the asymptotic region, $g'=g+1$.}

\section{Teichm\"{u}ller parameters for boundary of Euclidean multiple BTZ black holes}
\label{sec:teich-multiBTZ}

We now attempt to induce complex structure (conformal structure) into the boundary of a multiple BTZ black hole spacetime. The biholomorphic class (conformal class) of the complex structure (conformal structure) will be parameterized using the Teichm\"{u}ller parameters.
One general way to derive the Teichm\"{u}ller parameters (or moduli parameters) 
is to determine the period matrix~\cite{FK}, 
which is defined using holomorphic 1-forms and a canonical homology basis. 
In general, however, it is difficult to determine the period matrix for a quotient 
space which is a higher genus Riemann surface from discrete subgroup 
$\Gamma'$ of the identification isometry, 
since an exact expression of its holomorphic 1-form (or automorphic form) is not known. 
Consequently, although in the case of a single BTZ black hole spacetime 
the Teichm\"{u}ller parameter for its boundary can be successfully 
determined from the discrete subgroup $\Gamma$ of the isometry 
as demonstrated in section \ref{sec:singleBTZ},  
it is not easy task to obtain an exact expression for the
Teichm\"{u}ller parameters for the boundary of a Euclidean multiple BTZ black hole spacetime.

To overcome this difficulty, we make the assumption that all of the black holes have a large separation from each other, and derive an approximate expression for the Teichm\"{u}ller parameters for the spacetime boundary.

\subsection{Period matrix and pinching parameter}
\label{sec:period-matrix}

The definition of the period matrix of a genus $g$ Riemann surface is~\cite{FK,Tuite99}
\begin{align}
\Omega_{ij}=\int_{B_j} \nu_i,\quad( i,j=1,...,g ),
\end{align}
where $\nu_i, (i=1,...g)$ are a set of $g$ holomorphic 1-forms normalized by
\begin{align}
\int_{A_i}\nu_j=\delta_{ij}
\end{align}
for a canonical homology basis $A_1,...,A_g, B_1,...,B_g$.
For a genus $g$ Riemann surface, 
the elements of the period matrix form a set of independent (complex) parameters 
with a positive definite imaginary part, 
so that the Teichm\"{u}ller space is ${\mathbb H}_g$, 
i.e., the $g$-dimensional Siegel complex upper half plane. 
Though this rigorous definition of the period matrix provides full 
information about the Teichm\"{u}ller space, 
in general it is difficult to obtain the 
exact form of the $g$ holomorphic 1-forms for $g\geq 2$ Riemann surfaces.

However, for a genus $g$ Riemann surface that includes a narrow bridge structure, 
it is possible to approximately determine the period matrix 
from two period matrices of Riemann surfaces with a lower genus,
which are sewn together.
Therefore, we consider only the case of two Riemann surfaces connected by a narrow bridge. 
A general method for calculating $\Omega$ for any two sewn-together Riemann 
surfaces was reported by Yamada~\cite{Yamada}.

\begin{figure}[tp]
\centering
\includegraphics[width=7cm,clip]{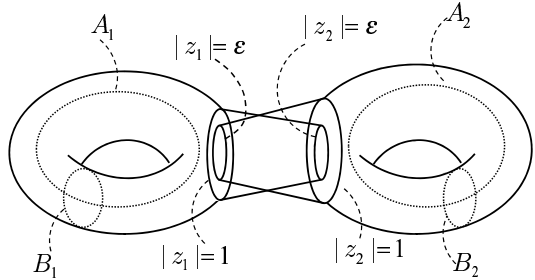}
\caption{Two tori are sewn together by identifying the annular regions 
$|\epsilon|\leq |z_a|\leq 1$ via the relation
$z_1 z_2 =\epsilon$. }
\label{fig:two-tori}
\end{figure}

In general, two compact Riemann surfaces $S_1$ and $S_2$ of genus $g_1$ and $g_2$, respectively,
can be sewn together, giving a Riemann surface of genus $g_1+g_2$.
This is achieved by using complex local coordinates $z_a$ on $S_a$ ($a=1,2$), 
and excising the two disks $|z_a|<|\epsilon|$, 
where $\epsilon$ is a complex parameter satisfying $|\epsilon|<1$.
The centers of the disks are taken to be at points $z_a =0$.
The two surfaces are sewn together by identifying the annular regions 
$|\epsilon|\leq |z_a|\leq 1$ via the relation
\[z_1 z_2 =\epsilon.\]

In the formulation given by Yamada~\cite{Yamada}, 
the period matrix can be expressed as 
a power series in $\epsilon$ as
\begin{align}
\Omega = \left(
\begin{array}{c|c}
\Omega_1 &  0 \\
\hline
0  &   \Omega_2
\end{array}
\right)
-2\pi i \epsilon
\left(
\begin{array}{c|c}
 0 &  {}^t\!W_1W_2\\
\hline %
   {}^t\!W_2W_1 &    0
\end{array}
\right)
+O(\epsilon^2),
\label{eqn:omg}
\end{align}
where 
$\Omega_a (\ a=1,2)$ are the period matrices for the genus $g_a$ Riemann surfaces, 
(if the Riemann surface is a torus, these correspond to the usual moduli parameters $\tau_a$).
The holomorphic functions $W_a$ are defined as
\begin{align}
W_a = (f_1(0),...,f_{g_a}(0)),
\end{align}
where $f_k$ are holomorphic functions appearing in the normalized holomorphic 1-form
at each set of local coordinates $z_a$,
$\nu_k=f_k(z_a)dz_a$ ($k=1,...,g_a$).
The contribution from the $\epsilon^2$ term in (\ref{eqn:omg}) yields the Weierstrass function, the standard elliptic Eisenstein series, and Bernoulli numbers. 
See \cite{MasonTuite} for further information.

The sewing process for two tori is illustrated in figure \ref{fig:two-tori}.
The identified annular regions can be thought of as a connecting bridge.
Note that as $\epsilon \rightarrow 0$, the connecting bridge is pinched down
and the Riemann surface degenerates into two tori 
with standard modular parameters $\tau_1$ and $\tau_2$.
The parameter $\epsilon$ is therefore referred to as the pinching parameter.
The parameters $\tau_1, \tau_2$ and $\epsilon$ form a set of moduli 
parameters for genus two Riemann surfaces.

The number of dimensions in the Teichm\"{u}ller space is determined as follows.
A genus $g\geq 2$ Riemann surface consists of $g$ tori 
connected by $g-1$ bridges. It is then trivial to conclude that the Teichm\"{u}ller 
space is spanned by the following parameters: 
$g$ moduli parameters $\tau_a\in\mathbb{C}, (a=1,2,...,g)$, 
$g-1$ pinching parameters $\epsilon_b\in\mathbb{C}, (b=1,2,...,g-1)$ and 
the relative positions of the bridges (symbolically, for example, 
$z_c\in\mathbb{C}, (c=3,...,g)$) which are also the degrees of freedom for Teichm\"{u}ller deformation. 
Therefore, the dimensions are given by
\[
g\cdot 2+(g-1)\cdot 2+(g-2)\cdot 2=6g-6,
\]
which reconfirms the well-known dimensions of the Teichm\"{u}ller space.

\subsection{Period matrix for boundary of E2-BTZ}
\label{sec:epsilon}
The multiple BTZ black hole geometry is constructed by repeating 
to bring an additional BTZ black hole to the single BTZ black hole 
geometry. At the spacetime boundary, this is achieved by connecting a 
torus to the Riemann surface while ensuring that the black holes remain 
far apart from each other. 
Then all that is necessary for understanding the complex structure of a multiple BTZ 
black hole geometry is reduced to introduce complex structure on the 
boundary of the double BTZ black hole geometry (E2-BTZ). 
For E2-BTZ, the lowest order quantities in (\ref{eqn:omg}), denoted by $\Omega_1$ 
and $\Omega_2$, are given by the 
moduli parameters for the first and second BTZ black holes, 
that is, $\tau_{\rm BTZ}$ defined in (\ref{eqn:teich}).
The essential task is then to determine the pinching parameter 
$\epsilon$ for E2-BTZ. 
For that purpose, we relate the pinching parameter to the coordinate transformation performed during the construction of E2-BTZ.

In section \ref{sec:multiBTZ}, 
we made a second identification of the surfaces $\phi=\pm\pi$
after a $y_p$ Lorentz boost with an angle $2\alpha$, obtaining the E2-BTZ spacetime.
We now consider a $y_p$ Lorentz boost with an angle $-\alpha$.
The center of the two BTZ black holes is then at the surface $y_p=0$.
On the boundary, this surface becomes a circle. 
In the remainder of this section, 
we consider only the Riemann surface on the boundary of E2-BTZ,
and think of this boundary as a Riemann surface constructed by sewing
parts of the boundaries of two Euclidean BTZ black holes around the circle $y_p=0$.
Focusing attention on a single Euclidean BTZ before sewing, 
we investigate the complex structure around the circle. 
Since the complex structure of the boundary of the second black hole is 
naturally given by the previous coordinates (in which the second 
identification is made), 
we again consider a $y_p$ Lorentz boost with an angle $\alpha$ and 
move the circle in the $y_p$ direction.
The coordinate system for the disk model then becomes
\begin{eqnarray}
x_p' = \frac{x_1}{1+x_0\cosh\alpha}, \quad
y_p' = \frac{x_0 \sinh\alpha}{1+x_0\cosh\alpha}, \quad
z_p' = \frac{x_3}{1+x_0\cosh\alpha}.
\end{eqnarray}
Then, the cut circle $y_p=0$ is mapped to
\begin{eqnarray}
y_p'=\tanh\alpha, \quad  
x_p^{'2}+z_p^{'2} = \frac{1}{\cosh^2\alpha}.
\end{eqnarray}
In the Poincar\'{e} coordinates $(X,Y,U)$, 
the circle is described as
\begin{eqnarray}
U=0, \quad X^2+(Y-\frac{\cosh\alpha}{\sinh\alpha})^2 = \frac{1}{\sinh^2\alpha}.
\end{eqnarray}
That is, the circle is 
centered at $X=0, Y=\coth\alpha$ and has a radius ${\rm csch} \alpha$,
on the boundary of the second Euclidean BTZ black hole.
We consider $Z=X+i(Y-\coth\alpha)$ as a local complex coordinate 
on the boundary.
If the parameter $\alpha$ is large enough,
the radius of the circle becomes smaller than 1, and
the circle can be thought of as the cut circle $|Z| = |\epsilon|$.
Since the period matrix is expressed as a power series in $\epsilon$, we consider the case where $\epsilon$ is small,
corresponding to the situation in which the two BTZ black holes 
are far apart due to a transformation with a large boost angle. 
Therefore, the expression for the period matrix (\ref{eqn:omg}) 
is relevant when the distance between the two BTZ black holes is large.

\begin{figure}[tp]
\centering
\includegraphics[width=9cm,clip]{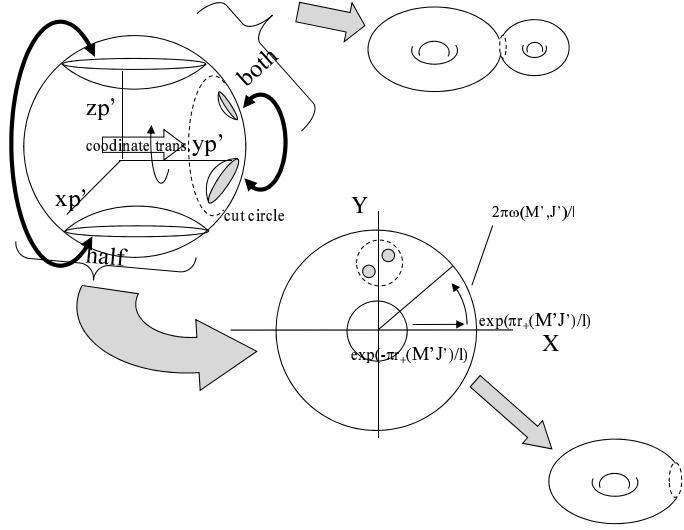}
\caption{The first and second identifications give a genus two solid 
 torus, corresponding to E2-BTZ. 
After a $y_p$ Lorentz boost with an angle $\alpha$, the cut circle 
 moves in the $y_p$ direction. The second identification in Poincar\'{e} 
 coordinates is shown in the lower figure, which gives a BTZ black hole with 
 a mass $M'$ and an angular momentum $J'$. In these coordinates, the cut circle is 
centered at $X=0, Y= \coth \alpha$, and has a radius of csch $\alpha$.}
\label{fig:E2-BTZ2}
\end{figure}

As described in section \ref{sec:multiBTZ}, the parameter $\theta$ allows a difference in orientation to exist between the two black holes. 
This corresponds to a rotation around the sewing annulus 
connecting the two boundaries of the Euclidean BTZ black holes. 
The pinching parameter can then be expressed as
\begin{eqnarray}
\epsilon={\rm csch}\alpha e^{2i\theta}.
\label{eq:epsilon}
\end{eqnarray}
That is, $|\epsilon|$ represents the radius of the cut circle and 
$\arg\epsilon$ is the twist angle of the connection.
Therefore, the pinching parameter for E2-BTZ has degrees on freedom 
corresponding to the relative positions and orientations of the two black holes.

Although in the above discussion, 
we considered the boundary of the second black hole,
much the same is true for the boundary of the first black hole
and we can introduce complex coordinates on it.
We identify the two sets of complex coordinates using the label $a=1,2$.
For each set of local coordinates, the pinching parameter can 
then be expressed using (\ref{eq:epsilon}).
In order to determine the period matrix up to first order in the power 
series in $\epsilon$, the holomorphic functions $W_a$ must be known for each torus.
Before sewing, the holomorphic 1-forms for the canonical homology basis $A_a,B_a$ for each torus on the boundary are 
\begin{eqnarray}
\frac{\D Z_a}{2\pi i (Z_a+i \coth\alpha)}, \quad ( a=1,2)
\end{eqnarray}
in the local complex coordinates $Z_a=X_a+i(Y_a-\coth\alpha)$ centered on the annulus.
By definition, the holomorphic functions for each torus are 
\begin{eqnarray}
W_a=\frac{-1}{2\pi \coth\alpha}.
\end{eqnarray}
This gives the period matrix to first order.

\section{Partition function for E2-BTZ in extremal CFT}
\label{sec:partition-function}

We will now show an application of the Teichm\"{u}ller parameters.
In general, the partition function for a boundary CFT 
is parameterized using the Teichm\"{u}ller parameters representing the conformal class of the boundary,
and here we consider the boundary CFT corresponding to pure AdS$_3$ gravity.

Witten~\cite{Witten:2007kt} 
argued that pure 3D gravity with a negative cosmological 
constant should be dual to a CFT on the boundary of a central 
charge $(c_L,c_R)=(24k, 24k)$, where $k$ is a positive integer.
This CFT factorizes into a holomorphic CFT and an anti-holomorphic CFT,
whose lowest dimensional primary field has a dimension $k+1$.
Such CFTs are referred to as extremal CFTs (ECFTs).
The genus one partition function for the ECFT, expressed as $Z_k(q)$, was determined by 
Witten~\cite{Witten:2007kt} to be a modular invariant one.
For $k=1$, the partition function is
\begin{eqnarray}
Z_1(q) &=& |J(q)|^2 = |j(q) - 744|^2, % \\
% Z_2(q) &=& |J(q)^2-393767|^2, \\
% Z_3(q) &=& |J(q)^3 --590651J(q)-64481279|^2, \\
% Z_4(q) &=& |J^4 -787535J^2-85975039J+7406924266|^2,
\label{eq:partition-function-genus1}
\end{eqnarray}
where $j=1728E_4^3/(E_4^3-E_6^2)$. $E_4$ and $E_6$ are the usual 
Eisenstein series of weights 4 and 6.
The parameter $q$ is given by $q=e^{2\pi i \tau}$, where 
$\tau$ is the moduli parameter for the thermal AdS$_3$ given 
by (\ref{eq:modular-transformation-AdS-BTZ}).
It is usual to express dual CFT partition functions in terms of $\tau$, 
rather than $\tau_{\rm BTZ}$.
Since the partition function (\ref{eq:partition-function-genus1}) is modular invariant,
it includes 
a contribution from the states corresponding to the thermal AdS$_3$ and the BTZ black hole.
As discussed in \cite{Witten:2007kt}, information about the BTZ black hole entropy can be extracted from the partition function.

Witten also showed that the partition function for an ECFT with $k=2$ can be 
uniquely determined on a hyperelliptic Riemann surface of any genus.
An explicit expression for a genus two partition function 
$Z_{k,g=2}(\Omega)$ was determined 
by Gaiotto and Yin~\cite{GaiottoYin} for $k=1,2,3$.
For $k=1$, it is
\begin{eqnarray}
Z_{k=1, g_2}(\Omega) &=& \left|\frac{C_1}{\chi_{10}}\left(
\frac{41}{4608}\psi_4^3 +\frac{31}{1152}\psi_6^2 - \frac{3813}{2048}\chi_{12}
\right) \right|^2, % \\
% Z_{k=2, g=2}(\Omega) &=& \bigg|\frac{C_2}{\chi_{10}}\bigg(
% \frac{574489}{12230590464}\psi_4^6
% +\frac{1125863}{1528823808}\psi_4^3\psi_6^2
% +\frac{159769}{764411904}\psi_6^4 \nonumber\\
%  && \qquad \quad -\frac{17809159}{905969664}\psi_4^3\chi_{12}
% -\frac{6550529}{226492416}\psi_6^2\chi_{12} 
% +\frac{91785533041}{154618822656}\chi_{12}^2 \nonumber  \\
% && \qquad \qquad \qquad -\frac{393767}{1572865}\psi_4^2\psi_6\chi_{10}
% + \frac{229938936071}{9663676416}\psi_4\chi_{10}^2
% \bigg)\bigg|^2
\label{eq:partition-function-genus-two}
\end{eqnarray}
where $C_1$ is a constant, $\psi_4$ and $\psi_6$ are Siegel modular 
forms of weight 4 and 6, respectively, and $\chi_{10}$ and $\chi_{12}$ are the cusp forms. 
For a detailed definition of these quantities, see \cite{GaiottoYin}. 
The expression (\ref{eq:partition-function-genus-two}) was also obtained by 
Tuite~\cite{Tuite99} in a different context.
For $k=1,2,3$, it was shown \cite{GaiottoYin} that
in the limit $\epsilon \to 0$, 
the partition function $Z_{k,g=2}(\Omega)$ factorizes as
\begin{eqnarray}
Z_{k,g=2}(\Omega) \to \frac{\rm const.}{|\epsilon|^{4k}}Z_k(q_1)Z_k(q_2),
\label{eq:Z-epsilon-limit}
\end{eqnarray}
where $q_a=e^{2\pi i \tau_a}$ and $\tau_a$ are the moduli parameters for each torus.

Gaiotto and Yin obtained the partition function (\ref{eq:partition-function-genus-two}) for a double torus, but its gravitational meaning was not discussed \cite{GaiottoYin}.
As shown in section \ref{sec:multiBTZ}, 
the boundary of E2-BTZ is topologically a double torus,
and E2-BTZ is a solution in pure AdS$_3$ gravity. 
Hence, the partition function (\ref{eq:partition-function-genus-two}) 
can be interpreted as being relevant to E2-BTZ. 
Since a small $\epsilon$ means a large separation between the two BTZ black holes, 
the factorization (\ref{eq:Z-epsilon-limit}) 
implies that the degrees of freedom for the two BTZ black holes 
become independent of each other in the separation limit. 
This factorization will be a general property for multiple BTZ black hole spacetimes. 
Though, for $g\ge3$, other parameters arise in 
addition to the pinching parameters $\epsilon_a$ and toroidal 
Teichm\"{u}ller parameters $\tau_a$, 
the partition function will also factorize into
partition functions for each individual BTZ black hole in the separation limit.

\section{Discussion and Conclusion} 
\label{sec:discussion}
The complex structure of the boundary of a Euclidean multiple BTZ black hole 
geometry was investigated, 
and was applied to Witten's conjecture concerning a pure AdS$_3$ 
gravity/ECFT correspondence.
For two widely separated BTZ black holes, 
we calculated the pinching parameter $\epsilon$
and obtained an approximate expression for the Teichm\"{u}ller parameters for E2-BTZ. 
It was found that the pinching parameter includes information 
about the positional relation between two BTZ black holes in the long-distance limit.
We also showed that the E2-BTZ boundary has a double-torus topology, and 
the genus two partition function for the boundary extremal CFT can be interpreted as being relevant to E2-BTZ.

The canonical ensemble entropy for BTZ black holes 
at fixed temperature and angular momentum 
can be obtained from the partition function
via the relation
\begin{eqnarray}
S= \frac{\partial}{\partial T}(T\ln Z).
\end{eqnarray}
The factorization then implies that, in the separation limit ($\epsilon \to 0$), the overall entropy of E2-BTZ is the sum of that for the individual black holes. 
In contrast, if the two BTZ black holes are not separated, this will not be the case, 
because the partition function is not factorized.
This is natural because a gravitational interaction is expected between the two black holes, 
leading to non-additive properties. 
Since the entropy for a single BTZ black hole is given by the Bekenstein-Hawking area law, 
this implies that the entropy of multiple black holes may not be the sum of the horizon areas~\cite{kurita-siino2}.

It is expected that thermodynamic quantities can be obtained from the partition function via thermodynamic relations.
The same holds true for black hole thermodynamics, for example as discussed by Gibbons and Hawking \cite{Gibbons:1976ue}. 
Therefore, it is natural that moduli parameters which describe the partition 
function are related to thermodynamic quantities.
Actually, the moduli parameter for the BTZ black hole geometry is expressed in terms of thermodynamic 
variables as given in (\ref{eq:tau-beta-Omega}).
Since for E2-BTZ, the pinching parameter $\epsilon$ also parameterizes 
the partition function in addition to $\tau_a$, 
it might be possible that $\epsilon$ is related to certain thermodynamic quantities, and this will be investigated in future work.

The above discussion is based on the assumption that the thermodynamics of 
a multiple black hole spacetime can actually be formulated.
To our knowledge, no such formulation has yet been developed.
However, from the viewpoint of the AdS/CFT correspondence, 
the existence of a multiple BTZ black hole solution in AdS gravity, and 
a partition function for the boundary CFT, 
implies that such a formulization is possible, at least for asymptotically AdS cases.

\section*{Acknowledgements}
We are grateful to 
H. Kodama for helpful comments.

%---------   References   ---------%

\end{document}